\documentclass[aip,jcp,reprint]{revtex4-1}
\pdfoutput=1
\usepackage{amsmath,amsfonts,amssymb,bm,graphicx,hyperref}
\usepackage{color}

\newcommand{\eps}{\varepsilon}
\newcommand{\ee}{\mathrm{e}}
\newcommand{\df}{d_{\rm f}}
\newcommand{\CC}{\mathcal{C}}

\begin{document}

\title{The melting of stable glasses is governed by nucleation-and-growth dynamics} 
\author{Robert L. Jack}
\affiliation{Department of Physics, University of Bath, Bath BA2 7AY, 
United Kingdom}
\author{Ludovic Berthier}
\affiliation{Laboratoire Charles Coulomb, UMR 5221 CNRS-Universit\'e de 
Montpellier, 34095 Montpellier, France}

\begin{abstract}
We discuss the microscopic mechanisms by which
low-temperature amorphous states, such as ultrastable glasses, transform
into equilibrium fluids, after a sudden temperature increase.
Experiments suggest that this process is similar to 
the melting of crystals, thus differing from
the behaviour found in ordinary glasses. 
We rationalize these observations using
the physical idea that the transformation process takes place 
very close to a `hidden' equilibrium first-order phase transition, which 
is observed in systems of coupled replicas. We illustrate our views 
using simulation results for a simple two-dimensional plaquette spin model,
which is known to exhibit a range of glassy behaviour. 
Our results suggest that nucleation-and-growth dynamics, as found
near ordinary first-order transitions, is also 
the correct theoretical framework to analyse the melting of ultrastable glasses.
Our approach provides a unified understanding of multiple 
experimental observations, such as propagating melting fronts,
large kinetic stability ratios, and `giant' dynamic lengthscales.  
\end{abstract}

\maketitle

\section{Introduction}

Recent experiments using vapor deposition methods have 
produced stable glass states 
with very low enthalpy~\cite{Swallen07,Kearns10,Dawson12,dalal,fromglen}, 
offering a new route for production of amorphous materials with 
controllable properties~\cite{facets}. Simultaneously, the discovery of 
such new amorphous materials raises exciting theoretical 
challenges~\cite{rmp}, because they open a new observational 
window on the behaviour of glassy materials. 

In particular, these novel glassy states are kinetically 
`ultrastable'~\cite{Swallen07,sepulveda}.  On heating
at constant rate, they recover back to equilibrium at a higher temperature 
than conventional glasses.
Alternatively, if the stable glasses are held at a fixed temperature 
above the glass transition, their
relaxation to equilibrium is much slower than that of conventional glasses.
Some stable glasses retain 
their glassy structure over periods up to 
$10^{5}$ times longer than the equilibrium structural relaxation time
of the equilibrium fluid~\cite{sepulveda}. 
In addition, the mechanism by which thin 
films of stable glasses transform into the 
equilibrium liquid appears strongly heterogeneous, 
accompanied by melting fronts that sweep through the system~\cite{swallen09}.
This process is 
reminiscent of the melting of crystalline materials, 
and is different from the behavior of ordinary glasses. For thicker films of vapor-deposited glasses,
the transformation mechanism is different again, and remains poorly understood. 
The crossover between thin-film and bulk behaviour defines a dynamic 
lengthscale characterizing the melting process, and 
experiments report a crossover length in the micrometer 
range~\cite{Kearns10}. Such a `giant' dynamic lengthscale  
is unexpected in supercooled liquids, in which 
the dynamic correlation lengthscales associated with 
equilibrium relaxation near the glass transition are typically 
a few nanometers~\cite{reviewmark,book}. 

These experimental observations remain poorly understood and 
are currently the subject of intense 
experimental investigations~\cite{Swallen07,Kearns10,Dawson12,dalal,fromglen,swallen09,sepulveda,2level14,beta15}.
They raise several interesting questions.  For example, what structural features are responsible for the stability of these materials? How do deposition conditions affect their properties? What is the microscopic mechanism for the recovery back to equilibrium of these stable states?  In this work, we concentrate on this last question, comparing the transformation kinetics 
of these amorphous materials with the melting of crystalline solids.  We argue that this process has a universal (material-independent) character, because of the presence of a nearby first-order phase transition~\cite{FranzParisi97,BerthierJack15}, with associated nucleation-and-growth phenomenology. The phase transition that we invoke to rationalise the observed behaviours takes place when two physical copies of the system are coupled to each other by a field $\eps$. This phase transition is therefore `hidden', because it cannot directly be accessed in experiments. However, we show that this theoretical construction is nevertheless extremely useful for understanding the physical dynamics of a single stable glass, as it transforms back into the liquid. 

To illustrate this theoretical picture, we use computer simulations of a simple spin model -- the triangular plaquette model (TPM).  This system does not capture the molecular details of supercooled liquids, but it does mimic many features of glassy materials, such as dynamical slowing down and spatially heterogenous dynamics, linked to growing dynamic and static correlation lengthscales~\cite{Garrahan02,Jack05caging,Jacklength,Jack06fdt,Cammarota12plaq,JackBerthier12}. In particular, the existence of growing static correlations in this model is accompanied by first-order phase transitions associated with coupled replicas~\cite{Garrahan14,Turner15,JackGarrahan16}, as also seen in molecular 
glass-formers~\cite{silvio99,giacomo10,Berthier13,BerthierJack15}. 
The idea that simple plaquette spin models of this type can be useful for describing glass-forming liquids is at the root of dynamical facilitation theory~\cite{Garrahan02,ChandlerAnnRev}. We show here that the TPM exhibits the universal features that we expect of stable glasses: kinetic stability, nucleation-and-growth phenomena associated with melting close to first-order phase transitions, and giant dynamic lengthscales.  Since these features are associated with a phase transition, we expect that results for this simple system also apply to atomistic models that have similar phase diagrams, and by extension, to experiments.

The TPM is particularly well-suited for the present study 
because it is relatively cheap to simulate computationally, compared 
with atomistic liquids. More importantly, a 
formidable advantage over off-lattice liquids is the possibility 
to prepare directly -- and {at no computational cost} --  
equilibrium configurations with arbitrarily low energy, 
without the need for simulating the vapor deposition process, 
or achieving brute force equilibration at low temperatures.  
By construction, therefore, our results can say nothing about the 
preparation of ultrastable glasses (this problem has been addressed 
computationally~\cite{depablo1,depablo2,dePablo14}), but we can 
shed light on their 
 behaviour upon sudden heating.
Other strategies  have been used to achieve similar effect, including 
a random pinning procedure~\cite{Hocky14ultra}, or simulations 
with kinetically constrained models~\cite{Leonard10}, which all 
permit to `plant'~\cite{Krz09} low-temperature configurations at no cost.

In comparing our results with those of kinetically contrained models (KCMs)~\cite{gst},
we note that while both plaquette models and KCMs are representative 
of dynamical facilitation theory, the KCMs 
do not undergo the thermodynamic phase transitions described here, because they are defined explicitly
as models of excitations (or defects) that lack any static interactions.
By contrast, the TPM is defined in terms of spin variables with simple local interactions -- the low temperature
behaviour of this model is characterised by long-ranged many-body spin correlations (amorphous order), 
as well as low energy excitations without static interactions, 
similar to those that appear in KCMs.   
The static many-body spin correlations in the TPM are essential for the analogy that we draw here with nucleation-and-growth.
Earlier simulations of atomistic liquids have invoked a similar
analogy with melting processes based on empirical
observation~\cite{Hocky14ultra}. 
Here, we show how to make these ideas concrete,
and, how they may be used to make quantitative predictions for the observed behaviour. 

The structure of the paper is as follows. 
Section~\ref{sec:theory} outlines our general theoretical setting, and Section~\ref{sec:model} describes the model that we consider.  Section~\ref{sec:transformation} describes the kinetics of the transformation process from stable glass back to equilibrium, and Section~\ref{sec:nucleation} investigates the mechanism of this process using spatio-temporal correlation functions.  In Section~\ref{sec:discuss} we discuss the main implications of our results, before concluding with a short outlook in Section~\ref{sec:outlook}.

\section{Theoretical background}
\label{sec:theory}

\subsection{Basic process: Bulk transformation 
of stable glasses}

\label{basic}

To describe our general theoretical setting, we use $\CC$ to denote a configuration of some glassy system (for example, this might represent the positions of $N$ particles within a liquid, or the states of $N$ spins in the TPM).  The potential energy of configuration $\CC$ is $E(\CC)$.  We prepare a stable glass state, which is associated with a probability distribution $P_{\rm st}$.
For example, we might take
\begin{equation}
P_{\rm st}(\CC) \propto {\rm e}^{-E(\CC) / T_0 },
\label{boltz}
\end{equation}
which corresponds to a thermal equilibrium 
distribution at some low temperature $T_0$. In Eq.~(\ref{boltz}) 
we have set the Boltzmann constant to unity.
In experiments performed with ultrastable glasses, 
thermalisation at low temperature is not guaranteed by the vapor deposition 
process, and the 
distribution $P_{\rm st}$ is not known.

At time $t=0$, we couple this initial configuration 
to a heat bath at temperature $T \geq T_0$ for which the 
average energy $\langle E \rangle_T$ is larger than its average in the stable glass state $\langle E \rangle_{\rm st}$.  
If the system has any kind of ideal glass transition then we also 
assume that $T$ is higher than this temperature.  
After some (possibly very long) time,
the system will recover back to equilibrium at temperature $T$.  
The time $\tau_{\rm rec}$ taken for this process quantifies the kinetic 
stability of the original state.  It is natural to measure 
this time relative to the equilibrium $\alpha$-relaxation 
time $\tau_{\rm eq}$ of the system measured at the same temperature $T$.
This suggests that the appropriate adimensional measure of the
kinetic stability of the glass is~\cite{sepulveda,Hocky14ultra} 
\begin{equation}
S = \frac{\tau_{\rm rec}}{\tau_{\rm eq}},
\end{equation}
which we call the kinetic stability ratio.  In experiments,  
$S = 10^3 - 10^5$. In previous simulations 
using off-lattice supercooled liquids, stability ratios of 
at most $S\approx 10^2$ were reported~\cite{Hocky14ultra,flenner}.

\subsection{Link with an equilibrium first-order transition
for coupled replicas}

\label{sec:eps-theory}

We now introduce the coupled replica setting originally devised by 
Franz and Parisi~\cite{FranzParisi97}.  They defined 
the overlap $Q(\CC,\CC')$ which measures the similarity between configurations $\CC$ and $\CC'$.  For identical configurations we have
$Q(\CC,\CC')=1$ while for independent random configurations one expects $Q(\CC,\CC')\approx 0$.  For a spin system, it is conventional to take $Q= \frac{1}{N}
\sum_i s_i s_i'$ where $s_i$ is the state of spin $i$ in configuration $\CC$ 
containing $N$ spins, and similarly $s_i'$ 
is the state of spin $i$ in configuration $\CC'$.  

For a fixed stable glass configuration $\CC_0$, we then consider a 
biased thermal distribution for configuration $\CC$ at temperature $T$:
\begin{equation}
P_{\eps}(\CC|\CC_0) \propto {\rm e}^{-[E(\CC)-\eps NQ(\CC,\CC_0)] / T}.
\label{equ:peps}
\end{equation}
Here, a positive value of the field $\eps$ 
biases the configuration $\CC$ to be similar 
to the reference configuration $\CC_0$.

\begin{figure}
\includegraphics[width=8.5cm]{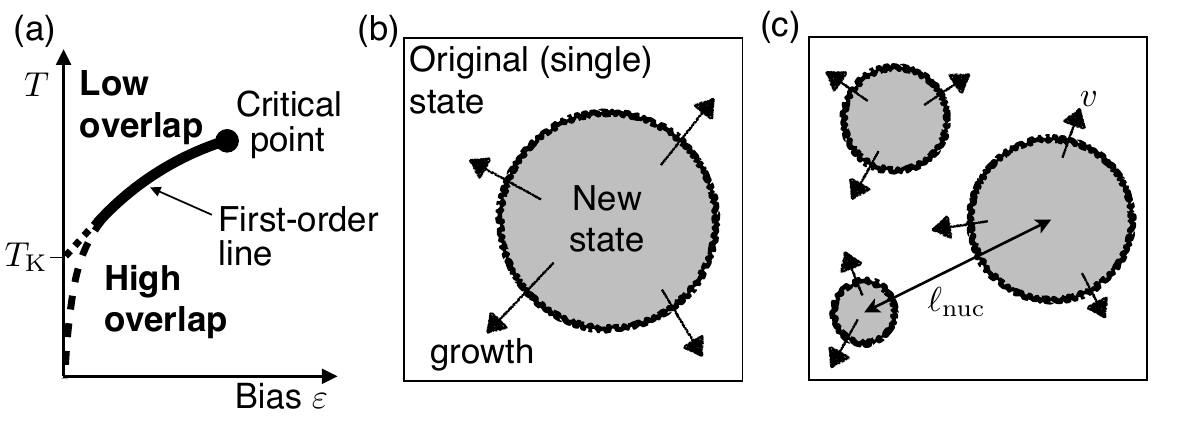}
\caption{(a) Phase diagram for coupled replicas.  There is a first-order phase boundary between high- and low-overlap phases.  Depending on the model, this phase boundary may intersect the $\eps=0$ axis at a finite temperature $T_{\rm K}$ (as happens in mean-field models~\cite{FranzParisi97}) or at $T=0$ (as in plaquette models~\cite{Garrahan14,Turner15,JackGarrahan16}).  (b)~Schematic figure illustrating nucleation and growth of a single droplet of a new state, within an original (reference) state. (c)~Schematic figure showing nucleation and growth in a large system, where multiple nuclei form and grow.  There is a large length scale $\ell_{\rm nuc}$ which is the typical spacing between nuclei.}
\label{fig:glass-nuc}
\end{figure}

If $\CC_0$ is a low temperature stable glass state and the temperature $T$ is not too high, one expects~\cite{FranzParisi97} a first-order phase transition to occur at some $\eps^*=\eps^*(T,T_0)$. The expected phase diagram 
in the plane $(\eps,T)$ is sketched in Fig.~\ref{fig:glass-nuc}(a).
At this transition, the average 
value of the overlap $\langle Q \rangle_\eps$ jumps from a small to a 
large value, as $\eps$ is increased through $\eps^*$.  For $\eps > \eps^*$, the configuration $\CC$ becomes trapped in the 
same metastable state as the reference configuration $\CC_0$.
We emphasize that the field $\eps$ is a 
thermodynamic quantity that appears directly in the energy function (\ref{equ:peps}), 
so that the transition at $\eps^*$ is an ordinary 
thermodynamic phase transition, not a non-equilibrium one.

What is the connection with the bulk melting of a single 
stable glass configuration? 
To see this, consider the following thought-experiment (or computer simulation). 
We generate a stable glass configuration $\CC_0$, 
and we initialise the system in this state by setting $\CC=\CC_0$.
Then, at time $t=0$, we connect the system to a thermal bath at temperature $T$, as in 
the usual setting of Sec.~\ref{basic}.
For $t>0$ we run the dynamics as usual, except that the system energy 
is now biased, as $E_\eps(\CC) = E(\CC)-\eps NQ(\CC,\CC_0)$, 
so that the system will eventually converge to the 
distribution function in Eq.~(\ref{equ:peps}).  
This distribution differs in general 
from an equilibrium state at temperature $T$, which
is recovered only for $\eps=0$. 

If we choose the field strength $\eps$ such that $\eps>\eps^*$ 
then  Eq.~(\ref{equ:peps}) means that $\CC$ will remain forever
in the same metastable state as $\CC_0$, so the overlap 
$Q(\CC,\CC_0)$ will remain close to unity. This implies that the 
system remains in a configuration close to the initial stable 
glass state for arbitrary long times: the glass never melts!  
If instead one has $\eps<\eps^*$, then the system will eventually relax to a 
state whose overlap with the initial glass configuration is low. In this case,
it should sample configurations similar to the equilibrium fluid 
at temperature $T$.
In other words, the field $\eps$ gives an additional handle to control
the kinetic stability ratio $S$ 
of the glass when heated to a temperature $T \geq T_0$.  The ratio $S$ 
can then be tuned from the physical value obtained 
at $\eps=0$, up to $S \to \infty$ when $\eps \to \eps^*$. 
We argue that this new handle, which {\it allows us to produce 
glasses with arbitrary-large kinetic stability ratio}, provides a key 
to a deeper understanding of the melting process and makes our study
experimentally relevant.

The central point of our paper 
is that the case $\eps < \eps^*$ includes the physical 
melting dynamics which occurs at $\eps=0$.  In this case, our thought experiment corresponds to
the natural (unbiased) dynamics of $\CC$, which is independent 
of $\CC_0$, except for the transient effect of 
this initial condition. The existence of the first-order transition 
at $\eps^*$ becomes physically relevant for the melting process 
when $\Delta\eps = (\eps^*-\eps)$ is small, because the system is then very close 
to a first-order phase boundary.  In this case, the system can be 
expected to relax 
into the low-overlap stable phase by a nucleation-and-growth mechanism.
The range of $\eps$ over which this condition applies is discussed in Sec.~\ref{sec:avrami} below.
The result is that if the critical field $\eps^*$ itself is sufficiently small, 
the natural melting process for  
stable glasses occurs close to this first-order phase boundary, so
the nucleation-and growth phenomenology should be at play. 
The qualitative difference between ordinary and ultrastable glasses 
is then very clear, as $\eps(T,T_0)$ for a given $T$ 
decreases rapidly as $T_0$ becomes smaller, implying that the 
melting of more stable glasses occurs closer to the phase boundary than 
the one of ordinary glasses. Note finally that this argument
is fully independent of the existence of a finite 
temperature ideal glass transition $T_{\rm K}$. 

\subsection{Transformation kinetics near first-order transitions}
\label{sec:avrami}

If the transformation of a stable glass into a liquid occurs
near a first-order phase transition, this immediately 
suggests that behaviour similar to ordinary 
first-order melting should be observed. We briefly summarize
the features of these processes 
that are relevant for the present situation.

\subsubsection{Classical nucleation theory} 

Close enough to the phase boundary, the transformation 
from configuration $\CC_0$ will take place via the nucleation of 
a droplet of the low-overlap phase, as depicted 
in Fig.~\ref{fig:glass-nuc}(b).
Applying classical nucleation theory (CNT) to this 
physical situation, we express the free energy to grow a 
droplet of the new (low-overlap) state of size $R$ 
inside the old (high-overlap) phase as~\cite{debenedetti,sear}
\begin{equation}
\Delta F(R) \approx \gamma R^{d-1} - \Delta \mu R^d,
\label{equ:cnt}
\end{equation}
where $\gamma$ is an interfacial energy cost and $\Delta \mu > 0$ 
the free energy difference between the two phases, 
which is expected to scale as 
the distance to the phase transition, $\Delta \mu \propto (\eps^*-\eps)$.  
Maximising $\Delta F$ then gives the size of the critical nucleus, 
$R^* \sim \gamma/\Delta \mu$, and of the free energy barrier to be crossed, 
$\Delta F^* \sim \gamma^d / \Delta \mu^{d-1}$.  
Both $R^*$ and $\Delta F^*$ 
diverge at the phase boundary where $\Delta \mu \to 0$.

In the coupled-replica system, this picture is slightly more complicated since the reference configuration $\CC_0$ enters the problem as a source of quenched disorder~\cite{BerthierJack15,silvio13,marco14}. 
Physically, 
this means that (i)~the system is no longer translationally invariant, 
so nucleation events might take place 
preferentially in particular regions of the system
where the free energy barrier is particularly low, 
and (ii)~there will be important 
sample-to-sample fluctuations of $\gamma$ and $\Delta \mu$, 
which means that these parameters will depend on the specific 
reference configuration $\CC_0$. While 
these two effects are certainly relevant for the melting 
of real stable glasses, we shall neglect them in the following.
Our strategy is to first obtain a robust general picture 
of the physical process, leaving for future 
work a more careful study of how quenched 
disorder affects the simple description offered here. 
This represents a significant, but certainly
worthwhile, additional effort.

\subsubsection{Avrami kinetics} 

We can use the phase diagram in Fig.~\ref{fig:glass-nuc}a to rationalise the giant length scale and the heterogeneous relaxation observed in experiments.  The idea is that when $\eps^*$ is small, then the natural dynamics of the system at $\eps=0$ still corresponds to the regime where $(\eps^*-\eps)$ is small and positive.  In this case the system is dominated by nucleation-and-growth, where large length scales and heterogeneous relaxation are expected.

To see this, let us recall the Avrami picture of nucleation 
kinetics~\cite{sear,avrami39}.  
In a large system, the nucleation rate per unit volume is 
\begin{equation}
k_{\rm nuc}\sim {\rm e}^{-\Delta F^*/T}.
\label{knuc}
\end{equation}
That is, starting from a system of volume $V$ that is entirely in the high-overlap phase, the droplets of the low-overlap phase appear at random positions in the system, with total rate $k_{\rm nuc}V$, as sketched in Fig.~\ref{fig:glass-nuc}(c). These droplets grow with a characteristic velocity $v$, until such time as they encounter each other and start to overlap.  Thus, paraphrasing Avrami's derivation~\cite{avrami39}, the fraction $f$ of material in the original (high-overlap) state evolves as
\begin{equation}
\frac{\partial f}{\partial t} = - f \cdot k_{\rm nuc} t \cdot c_d v^d t^{d-1}
\label{equ:ft}
\end{equation}
where $c_d$ is a dimensionless 
constant that depends only on the spatial dimension, 
such that the factor $c_d v^d t^{d-1}$ is the mean rate of growth of new material due to a single droplet whose radius is randomly (uniformly) distributed between $0$ and $vt$.  The factor $k_{\rm nuc} t$ is the number of nucleation events that have occurred up to time $t$, and the factor of $f$ takes care of the fact that if new material is generated in a place where the system has already transformed then this has no effect on the amount of the old phase that remains. 
The resulting time dependence is
\begin{equation}
f_{\rm avr }(t) = \ee^{-(t/\tau_{\rm avr})^{d+1}},
\label{equ:avrami}
\end{equation}
 where the characteristic time for formation of the new phase is 
\begin{equation}
\tau_{\rm rec} = \tau_{\rm avr} \simeq \left(k_{\rm nuc} v^d\right)^{-1/(d+1)}.
\label{taurec}
\end{equation}  
The characteristic 
compressed exponential shape of the relaxation function in 
Eq.~(\ref{equ:avrami}) appears because droplets grow with a fixed velocity, 
so the rate of production of the new phase increases with time and is 
proportional to the surface area of these droplets.
The transformation time $\tau_{\rm rec}$ in Eq.~(\ref{taurec}) has a strong 
dependence on both $T$ and $T_0$ as it involves both the velocity $v$ of the 
front propagation (which presumably decreases rapidly as $T$ is decreased), 
and the nucleation rate $k_{\rm nuc}$, which varies exponentially 
with control parameters, see Eq.~(\ref{knuc}).

Note also that if quenched disorder in the system leads to heterogeneous nucleation, the factor $k_{\rm nuc} t$ in (\ref{equ:ft}) will only be linear in time for small $t$, and will cross over to a sublinear increase for larger times.  This may lead to an apparent reduction of the exponent $d+1$ that appears in the compressed exponential in (\ref{equ:avrami}), as found in 
Ref.~\onlinecite{castro03}.

\subsubsection{Emergence of a `giant' dynamic lengthscale}

There is an important length scale associated with this process, which is much larger than the size of the critical nucleus $R^*$ discussed above.  
The physical picture is that phase transformation of a large system involves many independent nucleation events, followed by growth of the resulting droplets of the new phase, until they coalesce.  This situation is sketched in
Fig.~\ref{fig:glass-nuc}(c).
The typical number of nucleation events that happen during the transformation is ${\cal N} \simeq k_{\rm nuc} V \tau_{\rm avr}$ so the typical distance between the independent nucleation events is 
\begin{equation} 
  \ell_{\rm nuc} = (V/{\cal N})^{1/d} \simeq (v/k_{\rm nuc})^{{1/(d+1)}}. 
  \label{equ:ell-nuc}
\end{equation}  
Near the phase boundary, the nucleation rate is extremely small,
$\log k_{\rm nuc} \sim -1/(\eps^*-\eps)$, whereas  
the velocity $v \sim (\eps^*-\eps)$ vanishes much more slowly.  
This means that $\ell_{\rm nuc}$ can become 
very large, or `giant', as it scales 
exponentially with the distance from the phase boundary,  
\begin{equation} 
\ell_{\rm nuc} { \sim \exp\left[ \frac{\Delta F^*}{T(d+1)} \right] \sim \exp \left( A/ |\eps-\eps|^\alpha \right), }
\label{lnuc}
\end{equation}
where $\Delta F^*$ is the free energy barrier within CNT, so 
$A$ is a constant that depends on the surface tension $\gamma$ between the phases 
and $\alpha$ is a constant (equal to $d-1$ within CNT).

The scale $\ell_{\rm nuc}$ appears as a sort of \emph{dynamical heterogeneity}
in the non-equilibrium transformation process.  It also leads to {\it strong finite-size effects} in the transformation kinetics. If the system size is less than $\ell_{\rm nuc}$ then the Avrami picture breaks down and the system transforms by a single nucleation event, followed by a droplet that quickly grows and takes over the system.  In this case the compressed exponential relaxation of 
Eq.~(\ref{equ:avrami}) is replaced by simple exponential relaxation associated with the waiting time for the first nucleation event to occur.  That is, for system sizes $L\gtrsim \ell_{\rm nuc}$, one expects relaxation to follow (\ref{equ:avrami}) but for $L\lesssim \ell_{\rm nuc}$ one expects instead
\begin{equation}
f(t) = {\rm e}^{-t/\tau_1},
\label{equ:exp}
\end{equation}
with $\tau_{\rm rec} = \tau_1 \sim 1/(k_{\rm nuc} V)$ 
the volume-dependent mean waiting time for the first nucleation event.
(Notice that this relation implies that smaller samples are more stable than larger ones.) Therefore, the exponentially diverging lengthscale $\ell_{\rm nuc}$
in Eq.~(\ref{lnuc}) is also the crossover lengthscale 
controlling finite-size effects.

We note that this picture, of nucleation followed by front propagation at finite velocity, requires two important assumptions.  First, it only makes sense if nucleation is rare enough (that is, $k_{\rm nuc}$ small enough) that the growing nuclei can be identified and observed before they start to overlap.  This condition can be interpreted as the finite-dimensional signature of a spinodal line -- roughly speaking, the spinodal is the point where the nucleation barrier is of the same order as the thermal energy, $\Delta F^*/T \approx 1$, so that nucleation is no longer rare, and the system becomes \emph{locally unstable} to phase transformation~\cite{debenedetti}.  
In terms of stable glass melting, this criterion sets an upper limit on $(\eps^*-\eps)$.  In addition, to observe Avrami-like nucleation-and-growth kinetics, one also requires that the growth velocity $v$ is large enough that nuclei of the new phase grow quickly once they are formed.  Equivalently, the length scale $\ell_{\rm nuc}$ should be much larger than the critical nucleus size $R^*$, since otherwise the arguments leading to (\ref{equ:avrami}) break down.  As $\eps\to\eps^*$, the critical nucleus $R^*$ diverges as a power law in $(\eps^*-\eps)$ while $\ell_{\rm nuc}$ diverges exponentially, so this condition is surely satisfied.  However, if this condition breaks down for smaller $\eps$ (including the case of unbiased dynamics, $\eps=0$), then one expects the transformation by nucleation-and-growth to be replaced by an alternative mechanism with different 
kinetics. This might be what happens in the melting of ordinary glasses.

\subsubsection{Relation to stable glass melting}

Assuming that the picture of Fig.~\ref{fig:glass-nuc} applies to stable glass melting, we arrive at the following predictions.  (i)~We expect Avrami kinetics 
as in Eq.~(\ref{equ:avrami}) for the transformation of large systems, with a crossover to simple exponential kinetics in smaller systems.  For systems close to phase boundaries, the length scale $\ell_{\rm nuc}$ associated with this crossover may become very large.  (ii)~The transformation process should be strongly heterogeneous, involving fronts moving with a typical velocity $v$, and dynamical correlations over length scales up to $\ell_{\rm nuc}$.  (iii)~If it is possible to introduce (in simulations) a bias $\eps$, length and time scales should grow rapidly as $\eps$ increases towards $\eps^*$. 

The first two of these predictions are consistent with observed 
experimental data~\cite{Kearns10,sepulveda,moreavrami}. 
In the following, we illustrate all three of these effects in the TPM.  
We also discuss some behaviour in this model that may be 
different from the experimental situation, and we discuss the reasons 
for these effects.

\section{The triangular plaquette model} 

\label{sec:model}

The TPM is defined on a two-dimensional triangular lattice~\cite{newman99,Garrahan02}.  In our computer simulations we use a rhombus-shaped system of $L^2=N$ sites, with periodic boundaries.  The spins are located on lattice sites and are denoted by $s_i=\pm1$ with $i=1\dots N$. We also identify upward-pointing triangular plaquettes on the lattice: each plaquette $\mu$ is associated with three spins $s_{i_\mu},s_{j_\mu},s_{k_\mu}$.  We define plaquette variables $n_\mu = (1-s_{i_\mu}s_{j_\mu}s_{k_\mu})/2$, with $n_\mu=0,1$. The energy of the system is 
\begin{align}
E & = -\frac{J}{2}\sum_\mu s_{i_\mu}s_{j_\mu}s_{k_\mu} \\ & = -NJ/2 + {J}\sum_\mu n_\mu
\end{align}
Hence plaquettes with $n_\mu=1$ are excitations (or excited plaquettes), which carry energy $J$.  

At equilibrium (and for large systems), excited plaquettes are distributed as an ideal gas, so the plaquette variables are  independently identically distributed with $\langle n_\mu \rangle = c = 1/(1+{\rm e}^{J/T})$.  In the following we fix the energy scale $J=1$, which also sets the temperature scale.  In finite periodic systems, it is convenient to take the size $L$ as an integer power of two, in which case thermodynamic properties of the model are free from finite-size effects~\cite{newman99}.  In this case, for any given configuration of the plaquette variables $n_\mu$, there is exactly one possible configuration of the spin variables $s_i$, which may be constructed directly~\cite{newman99,Turner15}.

The model evolves in time by flipping spins according to Metropolis rates: spin $i$ flips with rate given by 
$\min(1,\ee^{\Delta E_i/T})$, where $\Delta E_i$ is the change in energy required to flip the spin.  The ensures that the system converges a Boltzmann distribution $p(\CC) \propto {\rm e}^{-E(\CC)/T}$.   The dynamical evolution is implemented using a continuous time Monte Carlo (MC) method~\cite{bkl}.

When considering coupled replicas, the overlap between configurations with spins $s_i$ and $s_i'$ is $Q=\frac{1}{N} \sum_i s_i s_i'$.  The distribution of initial (stable glass) states is $p_{\rm st}(\CC) \propto {\rm e}^{-E(\CC)/T_0}$ with $T_0<T$.  For $T_0=0$, this means that the initial state always has all spins with $s_i=+1$, since this is the ground state of the model, which is unique since we  take periodic boundaries and the system size is an integer power of two.

Since the model is defined in two spatial dimensions, the phase diagram in Fig.~\ref{fig:glass-nuc}(a) applies only for the special case $T_0=0$.  The first-order phase transition meets the $\eps=0$ axis at $T=0$, since the thermodynamic properties of the model for $\eps=0$ are trivial for all $T>0$ (the system maps to an ideal gas of excited plaquettes).  The critical temperature in Fig.~\ref{fig:glass-nuc}a is then $T_c=0.38$ (see Ref.~\onlinecite{Turner15}).  For $T_0>0$, the phase transitions in Fig.~\ref{fig:glass-nuc} are destroyed by the quenched disorder that enters the problem through the random configuration $\CC_0$.  
In this case, the first-order transition in Fig.~\ref{fig:glass-nuc} is replaced by a smooth crossover~\cite{imry75,aiz89}, but the behaviour near this crossover can still 
resemble what happens near a phase transition: 
this effect will be demonstrated below. To observe a phase transition using a finite preparation temperature 
$T_0 > 0$, one should study a three-dimensional generalisation of the model~\cite{Turner15},
which would then allow detailed theoretical 
analysis of the effect of the quenched disorder on the melting dynamics. 

\section{Kinetics of stable glass melting}

\label{sec:transformation}

\subsection{Bulk melting in large systems}

\begin{figure}
\includegraphics[width=7.5cm]{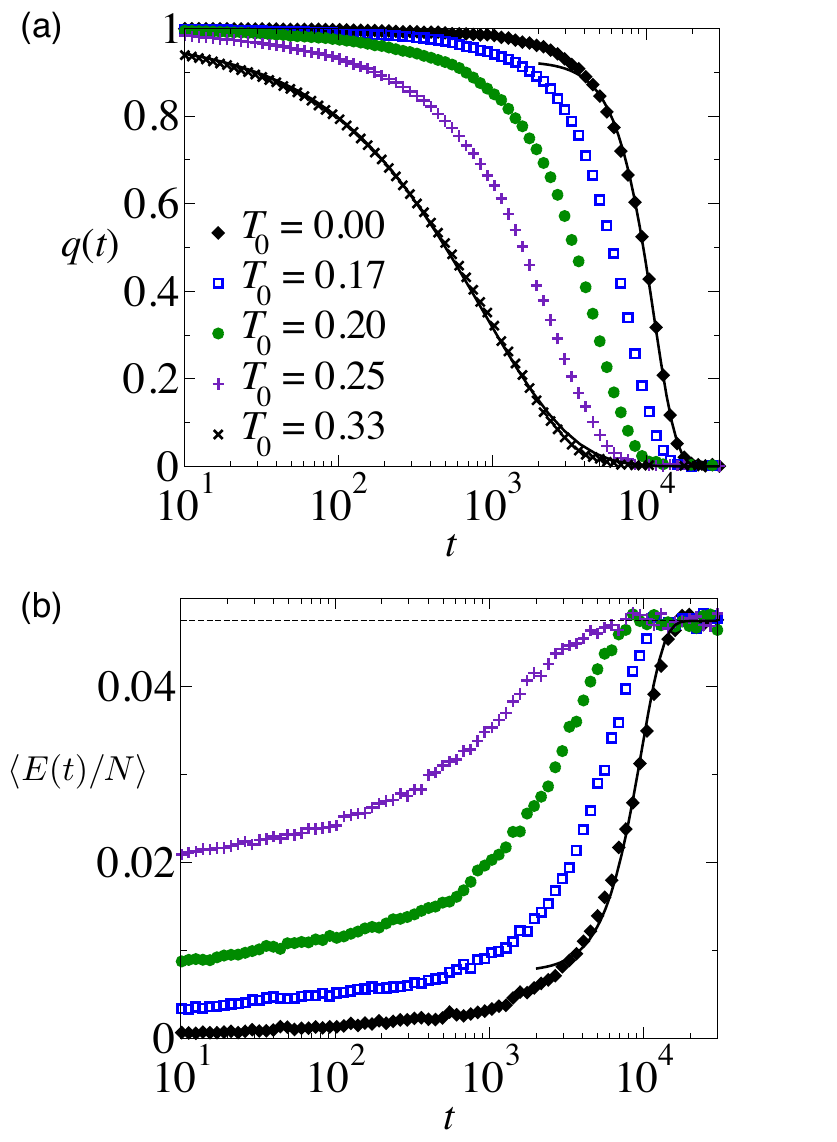}
\caption{Overlap $q(t)$ for stable glass recovery at $T=1/3$, varying $T_0$.  (Temperatures are quoted to 2 significant
figures throughout, the precise values used were $T_0=0,\frac16,\frac15,\frac14,\frac13$.)  Points are simulation results and lines are fits: 
for $T_0=0$ the late-time relaxation ($t\geq2000$) is fitted to an Avrami form
$q(t) = a \ee^{-(t/\tau)^3 }$. For $T_0=T$ (equilibrium relaxation) the fit is a stretched exponential $q(t) = a \ee^{-(t/\tau)^\alpha }$ with fitted $\alpha=0.74$.  The system size is $L=64$, which is large enough that the behaviour is representative of the limit $L\to\infty$.
(b)~Energy per spin, $\langle E(t)/N \rangle$, for the same process.  The dashed line
is the equilibrium energy $\langle E \rangle_{T}=N(1+\ee^{1/T})^{-1}$, and the
results for $T_0=0$ have been fitted with
an Avrami form $E(t) = \langle E \rangle_{T} - a \ee^{-(t/\tau)^3 }$.  
}
\label{fig:melt-vary-T0}
\end{figure}

As described above, we initialise a TPM in an equilibrium configuration at temperature $T_0$ and time $t=0$.  The system then evolves for $t>0$ by MC dynamics at temperature $T$, and eventually equilibrates at that temperature.  For fixed $T=\frac13$, Fig.~\ref{fig:melt-vary-T0} shows the time-dependence of this process for various $T_0$, through the time-dependent average overlap $q(t) = \langle Q(\CC_t,\CC_0) \rangle$ and the average energy per spin $\langle E(t)/N \rangle$.  The system size is $L=64$, which is large enough that these results are representative of the large-$L$ limit (for this specific example). 
Finite-size effects will be discussed in more detail below. 

For $T_0=0$ the initial configuration has all spins up.  Both the overlap and the energy are fitted in the long-time regime by Avrami (compressed exponential) form given in Eq.~(\ref{equ:avrami}), with $\tau_{\rm avr}=1.1\times 10^4$. At very early times, there are small fluctuations within the stable glass state that reduce $Q$ and increase $E$ -- these are not fitted by the Avrami form, which describes only the nucleation-and-growth process.  For this reason the fitting function is $q(t) = af_{\rm avr}(t)$ with $f_{\rm avr}(t)$ given by (\ref{equ:avrami}) and $a=0.925$ a fitting parameter.

Another special situation is when $T_0=T=\frac13$ in which case the average energy does not depend on time, by definition, and the overlap shows the equilibrium relaxation of the TPM.  In this case the overlap has a stretched exponential form, as is typical in glassy systems at equilbrium.  We show a fit to  $a\exp[-(t/\tau_{\rm eq})^\alpha]$ with $\tau_{\rm eq}=857$, $\alpha=0.74$ and $a=0.978$: note this is a three-parameter fit, in contrast to the two-parameter Avrami fit shown for $T_0=0$ where the compression exponent is fixed by theory. As $T_0$ increases from $0$ to $T$, the system crosses over from compressed exponential (Avrami-like) kinetics, indicative of nucleation-and-growth, to stretched exponential (glassy) kinetics, indicative of heterogeneous relaxation with a broad range of time scales. In the language of Fig.~\ref{fig:glass-nuc} this crossover takes 
place because increasing $T_0$ moves the relaxation dynamics at $\eps=0$ 
further away from the first-order phase boundary until its influence 
is no longer felt when $T_0 = T$.  

\subsection{Kinetic stability ratio}

\begin{figure}
\includegraphics[width=6.5cm]{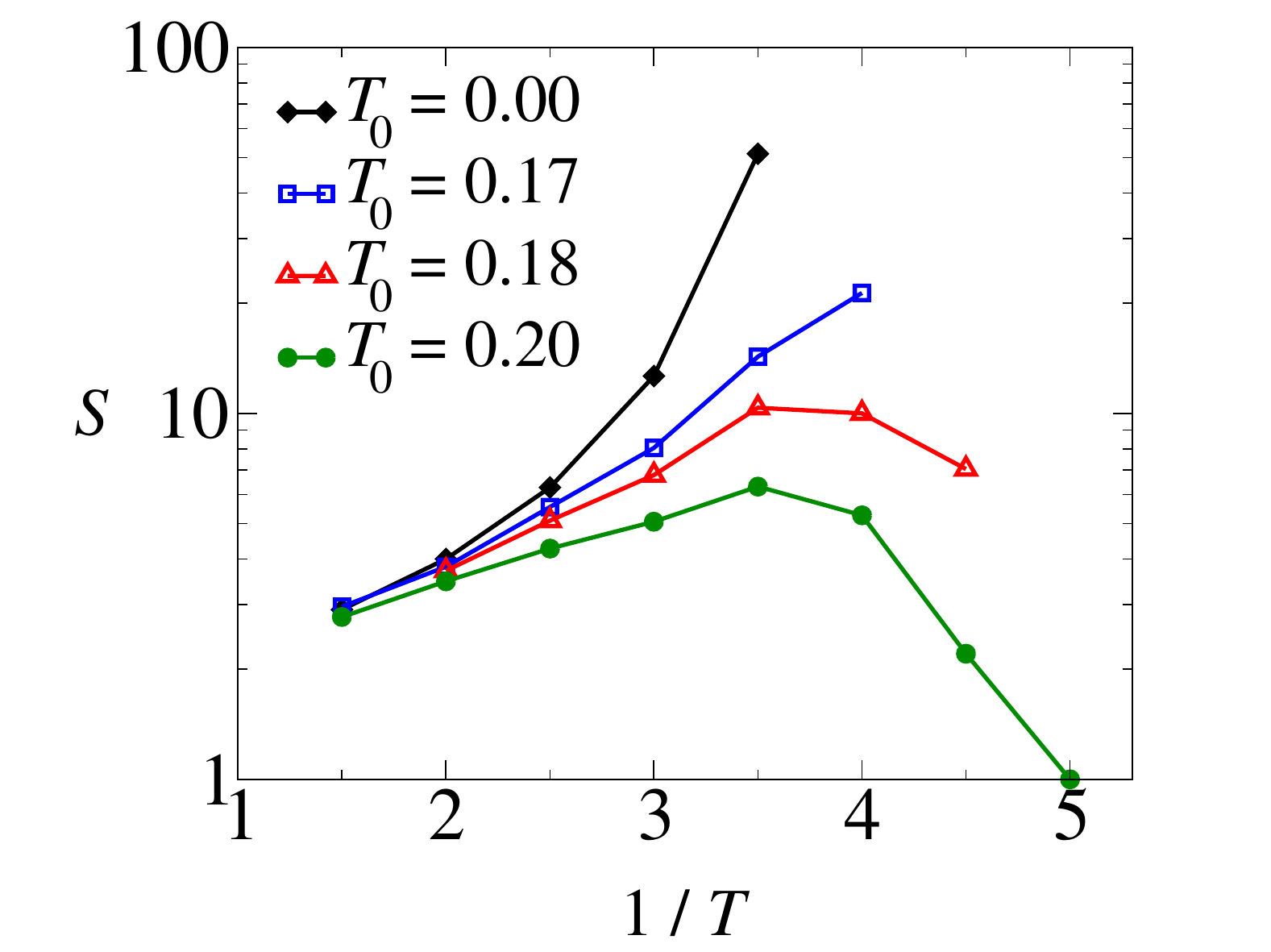}
\caption{Stability ratio $S=\tau_{\rm rec} / \tau_{\rm eq}$ as a function 
of inverse temperature $T^{-1}$ where melting occurs, 
for various preparation temperature $T_0$.}
\label{fig:stab-ratio}
\end{figure}

It is clear that the stable glass state with $T_0=0$ requires a long time to recover to equilibrium, compared with equilibrium relaxation at $T=\frac13$.  We extract the time for recovery to equilibrum as $q(\tau_{\rm rec})=1/{\rm e}$, and we identify $S=\tau_{\rm rec} / \tau_{\rm eq}$ as a stability ratio. 
We measure $S$ for various pairs $(T_0,T)$ and report our results in  
Fig.~\ref{fig:stab-ratio}. These results depend both on the stable glass state itself (through the temperature $T_0$) and on the transformation temperature $T$.  For a fixed melting temperature $T$, lower energy stable glasses are always more stable, as might be physically expected, but the dependence on the transformation temperature is non-monotonic.  Large stability ratios appear in a range of intermediate transformation temperatures $T$.

To understand this last result, note that for very high $T$, the rate for any spin to flip in the TPM approaches $1$, and all glassy behavior is lost. Hence $\tau_{\rm rec} \simeq \tau_{\rm eqm}\simeq 1$, so that when the melting temperature belongs to the non-glassy high-temperature regime, one necessarily has 
$S \approx 1$.  Physically, this effect may be attributed to the MC dynamics of the system, which implies that all spins are directly coupled to a 
stochastic heat bath, and this coupling is strong enough to melt the glass locally, without requiring any collective dynamics. Another trivial limit, on the 
other hand, is for $T=T_0$ where the recovery time extracted from the time-dependent overlap $\tau_{\rm rec}$ is equal, by definition, to the equilibrium relaxation time $\tau_{\rm eqm}$.  Hence one must again have $S=1$ at $T=T_0$. Therefore, for a given low $T_0$ value, $S\approx 1$ both at very high $T$ and when $T$ approaches $T_0$: the stable glass behavior becomes apparent only for 
intermediate $T$ values, which results in a non-monotonic 
temperature dependence. The only exception is when $T_0=0$ in which case 
we expect $S$ to increase monotonically on reducing $T$ without turning down 
again, because $\tau_{\rm eq} \to \infty$ as $T_0 \to 0$ and the position
of the maximum of $S$ has shifted to $T = 0$. 

We note that kinetic stability ratios $S$ found in experiments are often much larger than the values shown here, and they also tend to increase with temperature $T$, which is the opposite trend to the data for $T_0=0$ 
in Fig.~\ref{fig:stab-ratio}. 
In comparing absolute values of $S$ with experiments, we note that the temperatures $T$ considered here are relatively high, in the sense that equilibrium relaxation in the TPM at temperature $T=\frac13$ is only 2-3 decades slower than high-temperature (liquid-like) relaxation times.  The stability ratio increases rapidly (faster than an Arrhenius-law) on reducing $T$ so we might easily imagine reaching much larger stability ratios if we were able to perform simulations on the very long time scales comparable with experiment. It is indeed hard to imagine having a `more stable' glass than a perfectly thermalised $T_0 = 0$ 
initial configuration. 

From our results, it is not so easy to rationalize the apparent experimental finding that stability ratios $S$ tend to increase with $T$ over a wide temperature range (and not just for $T$ close to $T_0$).  
However, we note that the nature of the coupling of the stable glass to the thermal bath is rather different in experiments, compared to this kind of model, where all spins are strongly and directly coupled to the heat bath.  As discussed above, we expect this strong coupling to lead to $S\approx 1$ at 
high temperatures. Experimentally, such a trivial effect has not been reported,
even after temperature jumps to 
relatively high-temperatures in the mode-coupling regime~\cite{javier2}.
Of course, 
in experiments, each molecule is not directly coupled to a stochastic 
heat bath, and the calorimetric process following a sudden temperature change 
is less trivial than in simulations:
this might explain the discrepancy with our results in this 
regime, which is anyway not very relevant.

\subsection{System size dependence 
of transformation dynamics}

\begin{figure}
\includegraphics[width=7.0cm]{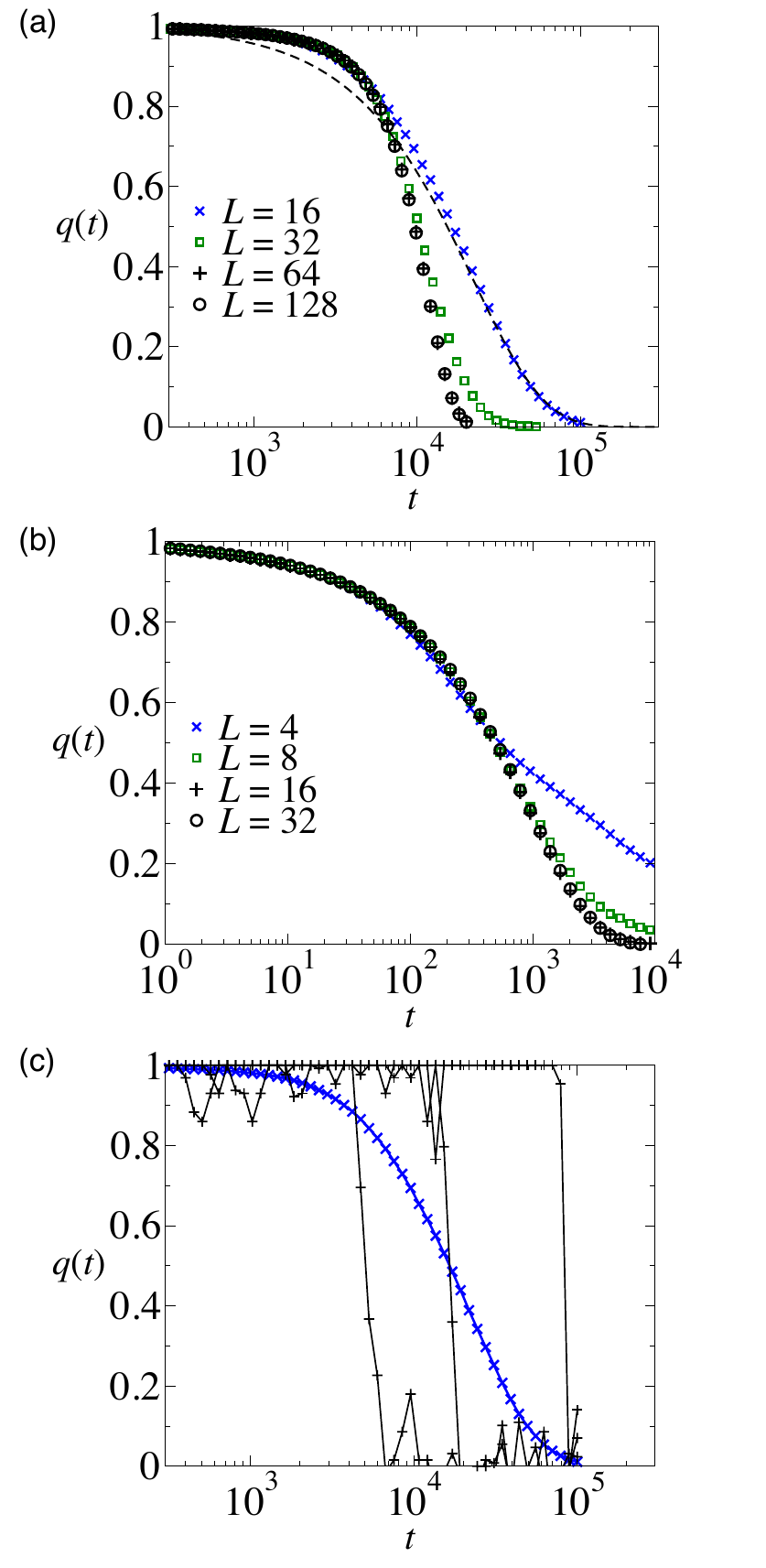}
\caption{(a) System size dependence of stable glass recovery at $(T,T_0)=(\frac13,0)$. 
 For large systems, one observes compressed exponential (Avrami-like) 
kinetics, as in Fig.~\ref{fig:melt-vary-T0}.  For smaller systems, the long-time
 relaxation is close to exponential (dashed line), 
 which we attribute to the exponentially-distributed waiting time for the first nucleation event.
(b)~Finite size effects for equilibrium relaxation at $T=T_0=\frac13$ are visible only for small systems $L=4,8$.
(c)~The average overlap during stable glass recovery at $L=16$ (data repeated from (a)), compared with three realisations of the time-dependent overlap $Q(\CC_0,\CC_t)$.  
In each individual trajectory, the system makes a rapid transformation between two states, involving a slow nucleation step followed by a very rapid growth of the new phase, which leads to an abrupt
decay of the overlap.}
\label{fig:melt-LL}
\end{figure}

As discussed in Sec.~\ref{sec:avrami}, the nucleation-and-growth picture of stable glass transformation predicts strong finite-size effects in the transformation kinetics. Figure~\ref{fig:melt-LL} shows this effect, for the case $T=\frac13$ and $T_0=0$ discussed above.  
Fig.~\ref{fig:melt-LL}(a) shows a significant finite-size effect in systems of linear sizes $L=32$ and $L=16$, whereas $L=64$ seems to have converged to the infinite system size limit.  This may be compared with the behavior shown in Fig.~\ref{fig:melt-LL}(b), which shows similar results for equilibrium relaxation at $T=\frac13$.  In this case, finite-size effects are significant only for $L=8$ and $L=4$.  This indicates that the non-equilibrium melting is characterised by a lengthscale that is of order four times larger than its equilibrium counterpart.
At equilibrium, the typical length scale for many-body spin correlations and dynamical heterogeneity in the TPM scales as $\xi \simeq {\rm e}^{1/(T d_{\rm f})}$ where $d_{\rm f}=\log_2(3)\approx 1.585$ is the fractal dimension of Sierpinski's triangle~\cite{Jack05caging}.  While the prefactor (proportionality constant) in the scaling relation for $\xi$ is not known, assuming that this factor is close to unity yields $\xi\approx 7$ for $T=\frac13$, consistent with Fig.~\ref{fig:melt-LL}(b).

Returning to the non-equilibrium relaxation of low-temperature initial states [Fig.~\ref{fig:melt-LL}(a)], the behaviour of $q(t)$ in the smaller system ($L=16$) is close to exponential, consistent with the theoretical prediction in Eq.~(\ref{equ:exp}) and in contrast to the compressed exponential found for Avrami kinetics in large systems.
Finally, Fig.~\ref{fig:melt-LL}(c) shows that while the average relaxation is exponential in a small system size, the individual trajectories relax with a simple two-state mechanism, where a single rare event leads to immediate transformation of the whole system.  The physical idea is that once nucleation has occured, the growth of the droplet of the new phase is so fast that it quickly takes over the whole system, so the system transforms by a single nucleation event. The exponential form is recovered by performing averages over different samples, because the instant of the melting fluctuates from one sample to another, presumably in a Poisson manner.

We again emphasise that while the length scale in Fig.~\ref{fig:melt-LL} 
is still relatively modest, and not comparable with the giant length scales observed in experiments, the stability ratio for this case is also relatively low ($S\approx 13$).  For lower transformation temperatures $T$, we expect much larger stability ratios accompanied by much larger lengthscales -- the difficulty is that the long time scales for these processes make simulations difficult.  In Section~\ref{sec:eps-results}, we show how this difficulty can be avoided by exploiting the coupled-replica construction described in Sec.~\ref{sec:eps-theory},
producing both large stability ratio and, indeed, giant dynamic lengthscales.

\subsection{Transformation dynamics for coupled replicas}
\label{sec:eps-results}

\begin{figure}
\includegraphics[width=8.5cm]{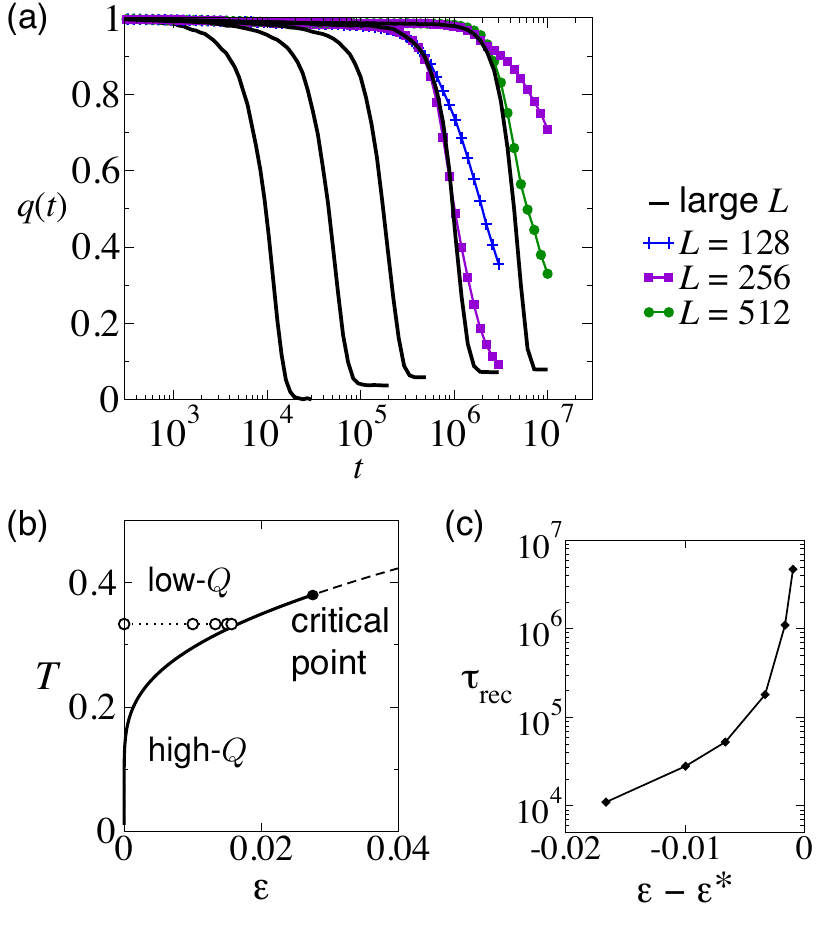}
\caption{(a) Overlap $q(t)$ showing stable glass recovery for $(T_0,T)=(0,\frac13)$, varying $\eps$ (increasing
left to right). The values of $\eps$ are $(0,0.01,0.0133,0.0150,0.0157)$, as indicated in panel (b) with 
open circles.
Solid lines show the behaviour that we find in the limit of large system size.  For the largest values of $\eps$, convergence of this limit requires system sizes of $L=512,1024$.
To illustrate these strong finite-size effects, numerical results for smaller systems are also shown.
(b) Phase behavior of a TPM, coupled by the field $\eps$ to a configuration $\CC_0$ at $T_0=0$.  The solid line indicates a first-order phase transition, which
separates high-overlap and low-overlap phases,  and ends at a critical 
point (black dot).  
The state points considered in (a) are indicated by open circles.
(c) The transformation times $\tau_{\rm rec}$ obtained from panel (a) grow rapidly as $\eps$ approaches the 
first-order transition, which occurs at $\eps^* \approx 0.0166$.}
\label{fig:melt-eps}
\end{figure}

The fits to Avrami theory in Fig.~\ref{fig:melt-vary-T0} indicate a nucleation-and-growth mechanism associated with a first-order phase transition.  We now show that this phase transition is the one anticipated by Franz and Parisi~\cite{FranzParisi97}, as discussed for the TPM in Refs.~\onlinecite{Garrahan14,Turner15}, and for a three-dimensional generalisation of this model in Ref.~\onlinecite{JackGarrahan16}.  

To this end, we now consider melting from $T_0$ to $T$ in the presence of a
positive biasing field $\eps > 0$, as 
discussed in Sec.~\ref{sec:eps-theory}. We 
show results in Fig.~\ref{fig:melt-eps}(a) for the transformation 
kinetics of a stable glass with $T_0=0$ at $T=\frac13$, 
as the biasing field $\eps$ is slowly increased.  Concentrating first on the bulk (large-system) behaviour,
one observes an increase of almost three orders of magnitude in the transformation time.  To rationalise this effect, Fig.~\ref{fig:melt-eps}(b) shows the phase diagram of the TPM in the presence of the coupling field $\eps$ for a reference temperature $T_0=0$. Note that since $T_0=0$, the reference configuration $\CC_0$ has $s_i=1$ for all $i$. Due to this simple reference configuration, the bias $\eps$ simply behaves as a magnetic field, and so this model belongs to the 
$2d$ Ising universality class~\cite{Turner15} and there is no quenched disorder, in contrast to cases with $T_0>0$. The first-order phase boundary is known exactly due to a duality symmetry of the model~\cite{Sasa10,Heringa89}, the position of the critical point was obtained numerically in Ref.~\onlinecite{Turner15} as $T_c\approx 0.38$. The path followed in Fig.~\ref{fig:melt-eps}(a) is represented in the phase diagram shown in Fig.~\ref{fig:melt-eps}(b), which explains the rapid 
growth of the transformation time $\tau_{\rm rec}$ as the 
transition is approached, as expected for first-order transitions.

The growth of the (bulk) transformation time $\tau_{\rm rec}$ with $\eps$ is shown
in Fig.~\ref{fig:melt-eps}(c), in a representation which clearly indicates
that it should diverge exponentially fast 
as $\eps \to \eps^* \approx 0.0166$. Because the temperature is constant in
this figure, the increase of $\tau_{\rm rec}$ translates into a 
an increase of the kinetic stability ratio $S$ from 
$S \approx 13$ at $\eps = 0$ to $S \approx 5500$ at $\eps = 0.0157$. 
The very large stability ratio reached near $\eps^*$ 
is comparable to the experimental measurements reported for ultrastable 
glasses, and we hypothesise that the melting process in both 
cases should be very similar. 

It is therefore experimentally relevant to demonstrate that such a large 
stability 
ratio is also associated with very strong finite-size effects in the transformation kinetics, as predicted in Sec.~\ref{sec:avrami}. These results are also shown in Fig.~\ref{fig:melt-eps}(a).  For the largest field considered ($\eps=0.0157$), there is a significant finite-size effect in melting dynamics even for $L=512$.  For $\eps=0.0150$, the behaviour for $L=512$ is consistent with the large-$L$ limit, but there is a significant finite-size effect for $L=256$.  Comparing with equilibrium relaxation at this temperature [Fig.~\ref{fig:melt-LL}(c)], the stability ratio of $S\approx 5500$ is accompanied by a giant length scale, in the sense  that it is around two orders of magnitude larger than the length scales characterising equilibrium behaviour of the simple liquid at the same 
temperature.

\begin{figure}
\includegraphics[width=8.5cm]{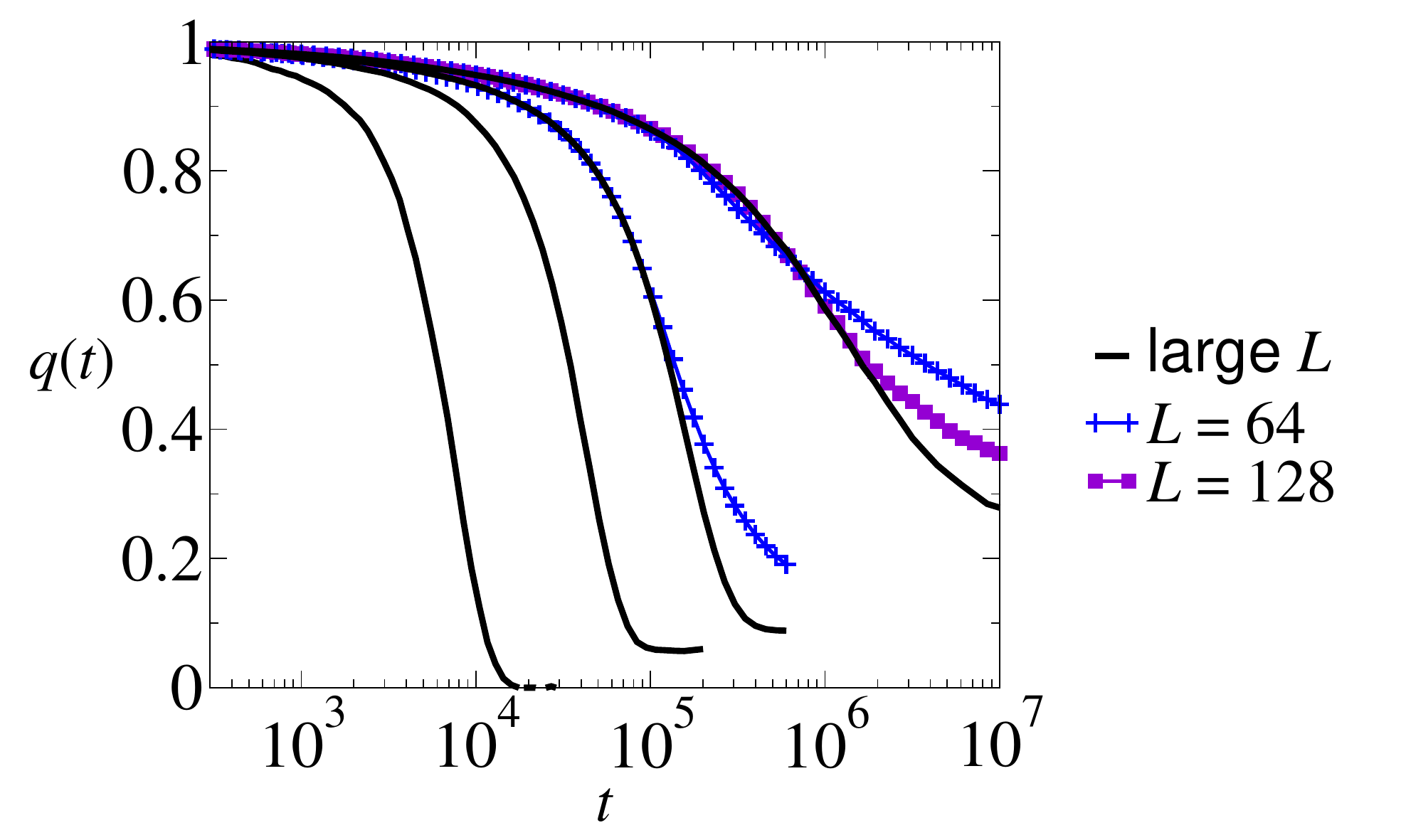}
\caption{Overlap $q(t)$ showing stable glass recovery at $T=\frac13$ and $T'=\frac16\approx 0.17$, 
for $\eps=0,0.0133,0.0167,0.0183$ (increasing from left to right).    
As in Fig.~\ref{fig:melt-eps}, solid lines show the behaviour in the large size limit, while symbols show data in smaller systems, to illustrate finite size effects.  For the largest $\eps$, we have verified that the large-$L$ limit is converged by comparing data for $L=256,512,1024$, which all agree to within statistical error (not shown).}
\label{fig:melt-bp6}
\end{figure}

For $T_0>0$ the fact that the TPM is a two-dimensional model means that the first-order phase transition shown in Fig.~\ref{fig:melt-eps}(b) is destroyed
by the quenched disorder that comes from the randomness contained in the 
finite temperature configuration $\CC_0$~\cite{imry75,aiz89}. (The same reasoning 
also explains the absence of a phase transition in the random field Ising model
in two dimensions.)
Nevertheless, one can still observe vestiges of this phase transition
on finite length and time scales.  Indeed, Fig.~\ref{fig:melt-vary-T0} shows that the stable glass transformation for $T_0>0$ is qualitatively very similar to that for $T_0=0$, at least when $T_0$ is low enough. 
To illustrate this effect more clearly, Fig.~\ref{fig:melt-bp6} shows results for $T_0=\frac16$, for increasing bias $\eps$.  For small fields $\eps$, these results resemble the ones in Fig.~\ref{fig:melt-eps}(a), with a transformation time that increases by nearly two orders of magnitude, with compressed exponential (Avrami-like) transformation kinetics.  We again attribute these results to nucleation-and-growth kinetics. As long as the critical nucleus is not too large, it is not apparent that the first-order phase transition has been destroyed by quenched disorder, since the effect of the disorder operates on large length 
scales~\cite{imry75,aiz89}.  However, as $\eps$ is increased, the critical nucleus grows and the effects of the quenched disorder become apparent as a change in transformation kinetics, crossing over from a compressed to a stretched exponential form.  
This shows that the effects of the (avoided) transition can still be felt, particularly when $\eps$ is not too close to $\eps^*$. 

In three dimensions, phase transitions survive~\cite{JackGarrahan16} for $T_0>0$, so one would expect nucleation-and-growth kinetics with a diverging time scale in that case too.  It would be interesting to investigate these effects in a three-dimensional model such as the square-pyramid model (SPyM), which is a three-dimensional generalisation of the TPM.  In particular, it would be useful to understand the influence of quenched disorder on nucleation and growth near the first-order transition in that case, but we postpone that investigation for a future study.  

\section{Nucleation-and-growth dynamics}

\label{sec:nucleation}

\subsection{Qualitative observations}

\begin{figure*}
\includegraphics[width=18cm]{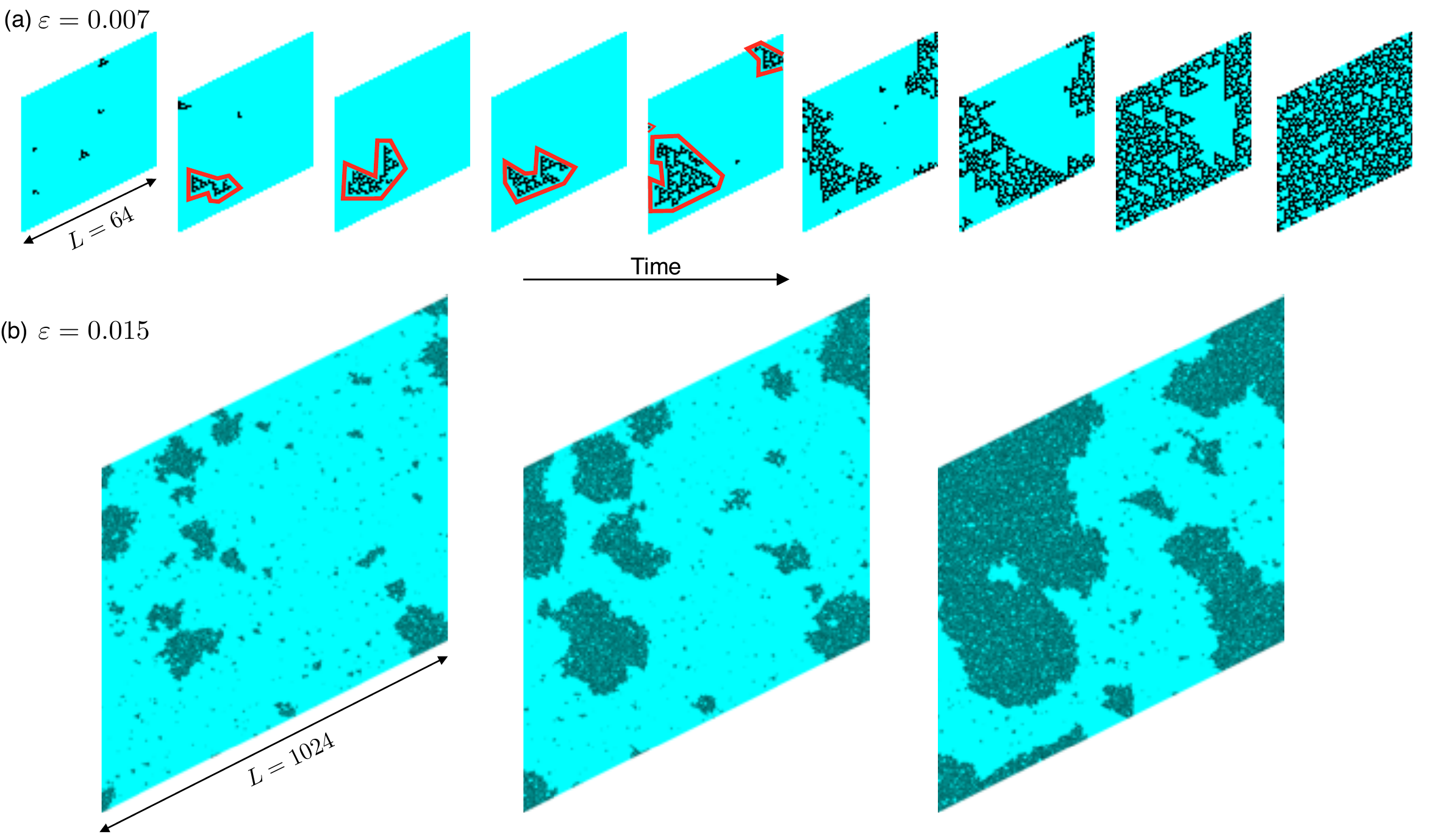}
\caption{Time-dependent local overlap $q_i(t)$ during nucleation 
and growth for $(T_0,T)=(0,\frac13)$.
Pale blue indicates sites where $q_i(t)=+1$, so $\CC_t$ matches 
the initial configuration $\CC_0$; black sites are where the configurations 
differ.  (a)~System size $L=64$ and $\eps=0.007$.
The configurations are equally spaced in log(time) between $0.15\tau_{\rm rec}$ and $1.06\tau_{\rm rec}$, with $\tau_{\rm rec}\approx 28000$.
Two independent growing nuclei of the low overlap phase 
are highlighted in red. 
At the final time, the system has reached equilibrium and approximately 
half of the spins match the initial condition, so the overlap is 
small ($q(t) \approx 0.02$). (b)~System size $L=1024$ and $\eps=0.015$ for 
times $(t/\tau_{\rm rec})=(0.36,0.52,0.75)$ with $\tau_{\rm rec}=1.1\times 10^6$.  There are multiple nucleation events, and the growing 
clusters merge and eventually percolate. At the merging time, the
`giant' lengthscale of the dynamic heterogeneity is about two orders of 
magnitude larger than in equilibrium at the same $T$.}
\label{fig:movie}
\end{figure*}

In this section, we show images of 
the heterogeneous nucleation-and-growth dynamics that takes place in the TPM as it transforms from an initial state at $T_0=0$ to an equilibrium state at temperature $T$.  To investigate this, we consider the local time-dependent overlap 
\begin{equation} 
q_i(t) = s_i(t) s_i(0), 
\end{equation}
which is equal to $+1$ if spin $i$ is in the same state as it was in the initial (reference) configuration $\CC_0$. Similar snapshots 
have been produced in earlier simulations of ultrastable glasses 
produced by random pinning~\cite{Hocky14ultra}.

We show in Fig.~\ref{fig:movie}(a) how the overlap evolves as a 
system transforms from a low energy initial condition to an equilibrium state,
for a representative trajectory at $T=\frac13$, $\eps=0.007$.
In that case, the system size is $L=64$. As time increases, we see the 
emergence of a first nucleation event (highlighted in red), followed 
by a second one at a later time (also highlighted). 
These two droplets then rapidly expand and fill the entire
system. At the time when the growing domains merge, the 
dynamic heterogeneity seems to be maximal, as the system is half relaxed in a spatially 
heterogeneous manner. At very long times, the system is homogeneous 
again, and resembles a typical equilibrium liquid configuration
at that same temperature. 
   
In Fig.~\ref{fig:movie}(b) we show a similar time series of spin configurations
for the same temperature $T=\frac{1}{3}$ but a larger field
value $\eps = 0.0150$, much closer to the transition point at $\eps^* \approx 
0.0166$.   
There is clearly a large length scale associated with this dynamical relaxation, which is accompanied by
the  much larger stability ratio shown in 
Fig.~\ref{fig:melt-eps}(c). 
To construct these images, we have used a system size $L=1024$.
The large length scale that is apparent in these
snapshots is consistent with Fig.~\ref{fig:melt-eps}(a) above, which showed that finite-size effects are significant for this process even for 
system sizes up to $L=256$.
We see multiple nucleation events, followed
by a rapid growth of the fluid phase invading the glass. 
These images provide a vivid visual demonstration of the melting 
process taking place in the present model.

\subsection{Dynamic lengthscales via four-point functions}

To analyse this behaviour quantitatively, we use the machinery of four-point correlation functions, which have been used extensively to discuss dynamical heterogeneity in glassy systems at equilibrium~\cite{book,cristina}.
Similar correlation functions were calculated for nucleation and 
growth processes~\cite{Krz11-1,Krz11-2}, and were measured 
also during the melting of randomly pinned glasses~\cite{Hocky14ultra}.

\begin{figure}
\includegraphics[width=6.5cm]{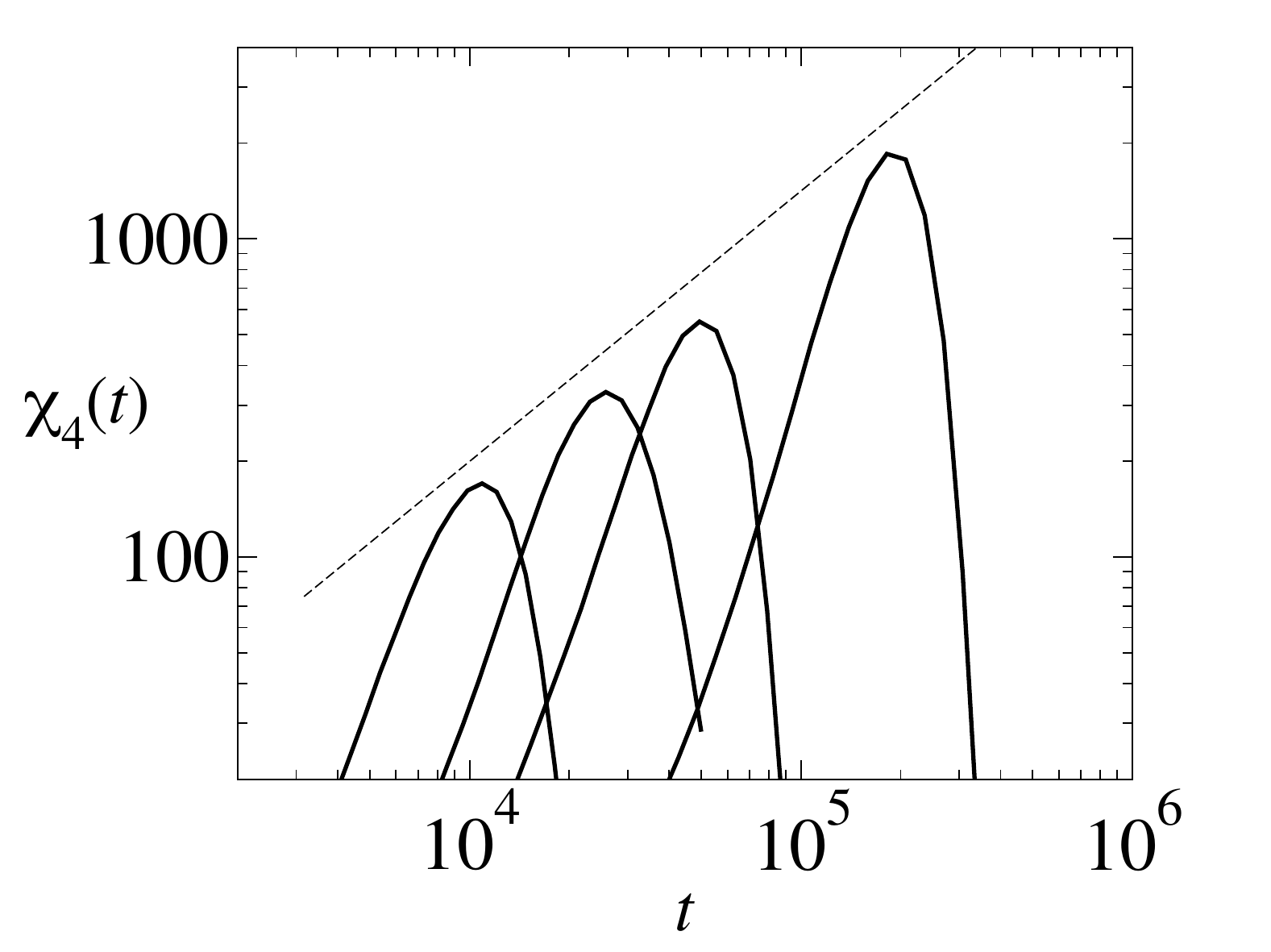}
\caption{Four-point susceptibility $\chi_4$ for $(T_0,T)=(0,\frac13)$ as $\eps$ is varied from $0$ to $0.013$ (increasing from left to right).  The increasing recovery time is accompanied by an increase  in dynamical heterogeneity.  The dashed line indicates $\chi_4 \sim t^{0.85}$, showing that the dynamic heterogeneity lengthscale increases algebraically with the 
kinetic stability of the glass.}
\label{fig:chi4}
\end{figure}

Four-point correlation functions are constructed from the overlap $q_i(t)$ as
\begin{equation}
g_{4,ij}(t) = \langle q_i(t) q_j(t) \rangle - q(t)^2.
\end{equation}
We emphasise that these averages run over both the random initial condition and the stochastic dynamics of the model.  This means that $g_{4,ij}$ depends only on the relative positions of sites $i$ and $j$, and that $\langle q_i(t) \rangle = N^{-1} \langle \sum_i q_i(t) \rangle = q(t)$.  

The function $g_{4,ij}$ measures the correlations of the overlap between sites so it characterises the correlated regions shown in the snapshots of 
Fig.~\ref{fig:movie}.  For a simpler characterisation of the strength of these correlations (or the size of the correlated domains), we also consider the four-point susceptibilty
\begin{align}
\chi_4(t) & = \left\langle \frac{1}{N} \left[\sum_i q_i(t)\right]^2  - N q(t)^2 \right\rangle
\\ & = \frac{1}{N} \sum_{ij} g_{4,ij}(t).
\end{align}

Fig.~\ref{fig:chi4} shows results for $\chi_4(t)$ for $T_0=0$, $T=\frac13$, and increasing $\eps$.  As expected for a system undergoing dynamically heterogeneous relaxation, the four-point susceptibility is non-monotonic in time, with a peak close to $\tau_{\rm rec}$, where $q(t)\approx 1/{\rm e}$.  The maximum value of $\chi_4$, which we denote by $\chi_4^*$, reflects the volume of domains of high (or low) overlap, as seen in Fig.~\ref{fig:movie}.  The significant result from Fig.~\ref{fig:chi4} is that $\chi_4^*$ increases strongly as $\eps$ is increased, providing a quantitative comparison of the increased heterogeneity associated with nucleation-and-growth as the phase boundary is approached.  We expect $\chi_4^*$ to be comparable with the maximal volume of correlated domains in Fig.~\ref{fig:movie}. Comparing with Fig.~\ref{fig:glass-nuc}, 
this size should be of order $\ell_{\rm nuc}^2$, 
which diverges exponentially fast as the phase boundary is approached.
Because the transformation time $\tau_{\rm rec}$ also diverges exponentially,
we expect a power law relation between $\chi_4^*$ and $\tau_{\rm rec}$, consistent with the simulation results in Fig.~\ref{fig:chi4}.
Such power law indicates a direct correlation between the stability 
ratio $S$ quantifying the kinetic stability to the relevant dynamic 
lengthscale controlling the melting process, as suggested 
before~\cite{Hocky14ultra}. 

\begin{figure}
\includegraphics[width=8.5cm]{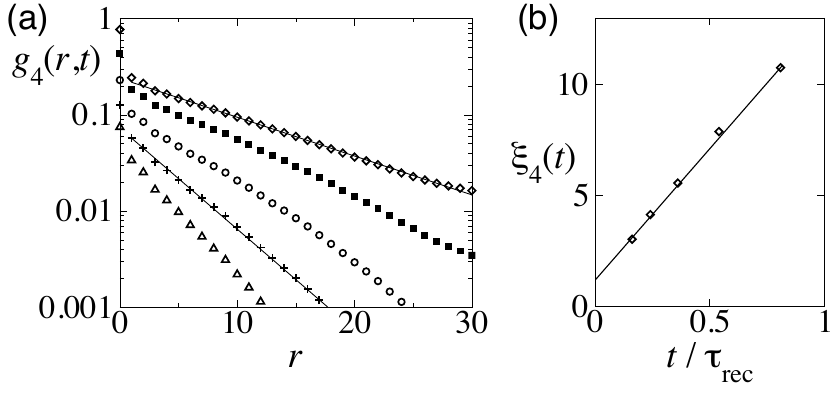}
\caption{Four-point correlation function $g_4(r,t)$ for $(T_0,T)=(0,\frac13)$ and $\eps=0$, for times $t/\tau_{\rm rec}=0.16,0.24,0.36,0.54,0.81$ (increasing from bottom to top). The lines (for selected times only) are fits to $g_4(r,t) = a(t)\ee^{-r/\xi_4(t)}$.  (b)~The time-dependence of the length $\xi_4(t)$ can be fitted as $\xi_4(t) = \xi_0 + vt$, indicating growth at a constant velocity.}
\label{fig:g4}
\end{figure}

To investigate this behaviour in more detail, we consider the four-point correlation function $g_{4,ij}(t)$. This function depends only on the relative positions of sites $i$ and $j$.  For simplicity, we take a circular average of this function, arriving at a function $g_4(r,t)$, where $r$ is the distance between sites $i$ and $j$. (There is fine structure in the dependence $g_{4,ij}$ on the orientation relative to the lattice of the vector connecting sites $i$ and $j$, but this is unimportant for the behavior considered here.)  For nucleation and growth, we expect domains to be compact, and hence
\begin{equation}
g_4(r,t) \simeq n(t) {\ee }^{-r/\xi_4(t) }
\label{equ:g4-exp}
\end{equation}
where $\xi_4(t)$ is the typical size of a growing domain of the new phase, and the prefactor $n(t)$ should be proportional to the number density of critical nuclei, at least in the early-time regime where droplets do not overlap.   Assuming as in Sec.~\ref{sec:avrami} that droplets of the new phase grow 
with velocity $v$, we expect
\begin{equation}
\xi_4(t) \sim vt + \xi_0. 
\label{equ:xi4-lin}
\end{equation}

Results for a representative state point (with $\eps=0$) are shown in Fig.~\ref{fig:g4}, including fits to Eqs.~(\ref{equ:g4-exp}, \ref{equ:xi4-lin}).  The
agreement is good, with a maximal domain size $\xi_4^* \approx 11$, consistent with the observation of Fig.~\ref{fig:melt-LL} that a system size $L=16$ is not large enough to recover bulk behaviour at this state point.  

We emphasise also that the linear growth with time of the (non-equilibrium) 
dynamic heterogeneity length in Eq.~(\ref{equ:xi4-lin}) again differs 
strongly from the subdiffusive behaviour found in 
equilibrium studies of dynamic heterogeneity~\cite{cristina}. 
This result shows that 
the propagation of mobility from rare nucleation sites is qualitatively
similar to the heterogeneous melting taking place from the interface in 
experimental work on ultrastable glasses, even though we observe 
the analog of `homogeneous' melting~\cite{sear}, 
i.e. nucleation initiated  
from the bulk rather than from an interface. For the present model, we expect 
the velocity $v$ to scale roughly as
\begin{equation}
\label{velocity}
v \simeq \frac{\xi_{\rm eq}}{\tau_{\rm eq}},
\end{equation}
where $\xi_{\rm eq}$ is 
the equilibrium correlation length (of order $\ee^{1/(Td_{\rm f})}$ as discussed above), and $\tau_{\rm eq}$ is the equilibrium relaxation time.
{The physical reasoning leading to Eq.~(\ref{velocity}) is that on the low-overlap side of the front, the system has a near-equilbrium structure, so its dynamics are equilibrium-like.  At equilibrium, regions of linear size $\xi_{\rm eq}$ take a time of order $\tau_{\rm eq}$ to equilibrate.  Hence the front moves through the system by successive equilibration of regions of  size $\xi_{\rm eq}$, each taking a time $\tau_{\rm eq}$, leading to Eq.~(\ref{velocity}); 
see also the discussion in Ref.~\onlinecite{Tylinski2015}.

This scaling relation indicates that the velocity should only
depend on the final temperature $T$, and should scale 
essentially as $1/\tau_{\rm eq}$, since the temperature 
dependence of $\xi_{\rm eq}$ is much weaker than that of $\tau_{\rm eq}$.
The scaling in Eq.~(\ref{velocity}) is very much 
consistent with experiments~\cite{velocity12}. 
The temperature dependence of $v(T)$ has also 
been addressed in the context of RFOT theory~\cite{Wolynes09,Wisit13,Wisit14}.
Together with the snapshots in Fig.~\ref{fig:movie}, the linear time
dependence of $\xi_4(t)$ is strong evidence that this system is exhibiting nucleation-and-growth behaviour, consistent with Avrami's theory.  We emphasise that this dynamical behaviour is taking place for the natural (unbiased, $\eps=0$) behavior, even if the only phase transitions that occurs in this model happen for finite bias $\eps$.  This is a sense in which \emph{avoided} phase transitions such as the one shown in Fig.~\ref{fig:glass-nuc} can still provide explanatory behaviour for the natural dynamical behaviour of glass-forming systems.

\section{Discussion\label{sec:discuss}}

\subsection{Connection to experimental results}

There are three principal aspects of our results for the TPM that are relevant for the mechanism of transformation of stable glasses in experiments. Since the
simulations use periodic boundaries, the relevant comparison is 
with bulk stable glasses, or thick films.

(1) The transformation of stable glasses in experiments has been observed to be similar to crystal melting.  The TPM reproduces this effect
and we have explained this phenomenology by reference to the 
first-order transition shown in Fig.~\ref{fig:melt-eps}.  
In two dimensions the transition is destroyed
for $T_0>0$, but its signature can still be seen in the transformation 
kinetics.  In the experimentally-relevant three-dimensional case, the
transition will survive for $T_0>0$ so the mapping to first-order phase transformation should remain precise, and our interpretation should hold.

(2) We predict the emergence of a `giant'
length scale $\ell_{\rm nuc}$ from Eq.~(\ref{equ:ell-nuc}) 
that is essentially the spacing between independent nucleation events.
This length scale diverges at the first-order transition 
between high-overlap (stable glass) and low-overlap (fluid) states, 
but this transition is present only for $\eps>0$.  
In general, the length scale is controlled by the nucleation 
rate $k_{\rm nuc}$, which depends strongly on the free energy 
difference $\Delta\mu$ between the stable glass and equilibrium fluid 
states.  For the physical case $\eps=0$, we expect
$\Delta \mu \approx T s_{\rm conf}(T) - u(T) + u(T_0)$ 
where $s_{\rm conf}$ is the configurational entropy density that is 
gained by the liquid during the transformation~\cite{daniele},
while $u(T) - u(T_0)$ is the increase in internal energy due to the
temperature difference.
At fixed $T$, more stable glasses have lower $u(T_0)$,
and these will therefore be associated with larger length scales.
(More strictly, $u$ would be a free energy that includes
the entropy associated with intra-state fluctuations and enthalpic 
contributions from volume changes during the tranformation process.)

Our interpretation of the giant length scales observed as finite-size 
effects in experiments is that for thin films, critical nuclei are so 
rare that propagating
fronts coming from the boundaries of the system can travel through the 
entire film before any nucleation event takes place.  The thick-film
(bulk) limit sets in only when homogeneous nucleation and growth has a 
significant effect on relaxation.  We expect this crossover to take
place at the point where the film thickness becomes comparable to the 
typical spacing between nucleation events. 
For a finite film of thickness $W$, 
one should compare the time for a mobility front to spread 
from the boundary through the system, $W/v$, with the 
time $\tau_{\rm avr}$
for homogeneous transformation given in Eq.~(\ref{taurec}).  
One finds that the homogeneous transformation mechanism operates only if 
$W\gtrsim \ell_{\rm nuc}$.
For this
reason, we identify the large length scale $\ell_{\rm nuc}$ with 
the giant crossover length scale measured in experiments~\cite{Kearns10}, which is characterised
through the dependence on the film thickness $W$.
Our results suggest that this large length scale could be observed 
directly in bulk stable materials, and would appear as a
giant dynamic heterogeneity lengthscale. 

(3) The length scales and stability ratios observed in this work 
are much smaller, for $\eps=0$, 
than those in experiments.  We attribute this to the
relatively high transformation temperatures $T$ considered here.  
As noted above, equilibrium relaxation at these temperatures 
is only 2-3 decades slower than relaxation at the onset of glassy dynamics, in contrast to the experimental case where
relaxation is typically studied close to the experimental glass temperature.
For the TPM, we expect that the stability ratio $S$ and the length scale 
$\ell_{\rm nuc}$ should increase
significantly as $T$ is reduced, and are likely to diverge as $T\to0$, 
taking always $T_0\ll T$, or perhaps more precisely 
$\tau_{\rm eq}(T_0) \gg \tau_{\rm eq}(T)$.  
For this reason, we expect that the modest
length and time scales that we have found in this work are due to our restriction to state points where computer simulations are tractable -- our theoretical arguments are applicable to the large length and time scales found in experiments.
We have supported this claim using the 
biasing field $\eps$ at constant $(T,T_0)$ to promote by more than
two orders of magnitude both the kinetic stability ratio and 
the dynamic lengthscale of the melting process.  

In addition to these three points, a potentially important factor that we have not considered here is the heterogeneous nature of the nucleation process, given that realistic initial conditions in experiments are not translational invariant (in contrast to the special case $T_0=0$ for the TPM).  This leads to the possibility that nucleation will occur preferentially at particular starting points within the stable glass phase.  It would be interesting to investigate this effect further, either in the TPM or in its three-dimensional generalisation 
(the SPyM).  We also note that in kinetically constrained models such as those considered in Ref.~\onlinecite{Leonard10}, the coupled replica construction does not induce any kind of phase transition.  In that case, one expects melting of stable glasses to start at pre-existing defects, or `soft spots', where spins are able to flip -- this situation may be related to the presence of preferred sites for nucleation-and-growth in the scenario considered here, but seems to differ in that there will be no slow nucleation step before relaxation starts.  This comparison also merits further investigation.

\subsection{Crossover between stable glass melting and equilibrium relaxation}

Before ending, we return to a question that arises 
from Fig.~\ref{fig:melt-vary-T0}(a): can the equilibrium relaxation itself
occurring for $T_0=T$ be explained
by a similar nucleation argument to the transformation of the stable glass?  
The shape of the relaxation function is different, but the general
RFOT-like description of Bouchaud and Biroli~\cite{Bouchaud2004} 
would seem applicable 
in both cases (see also Ref.~\onlinecite{Jack05caging}).
We offer a scaling argument as to how these two regimes might be 
smoothly connected.

Starting with transformation from a stable glass with $T_0=0$, the usual CNT predicts that the free energy cost for a droplet of the new phase is
\begin{equation}
\Delta F \approx \gamma R^{d-1} - \Delta \mu R^d
\label{equ:cnt2}
\end{equation}
with a critical nucleus $R^* \sim \gamma/\Delta \mu$ and a barrier $F^* \sim \gamma^d / \Delta \mu^{d-1}$.  Both diverge at the phase boundary where $\Delta \mu\to0$.

For equilibrium relaxation, we imagine that the phase boundary is still present (as would be the case in $d=3$). 
However, the relevant state for equilibrium relaxation is much further from the phase boundary and so the critical nucleus is much smaller.
Also, the form of the `droplets' that mediate relaxation at equilibrium is different -- the droplets are fractal objects of size $R$ that contain
$N_{\rm drop} \sim R^{\df}$ spins and have an energetic cost that scales as $J\log_2 R$.  (For the TPM, $\df=\log_2 3$ is the fractal dimension of Pascal's triangle; for
the three-dimensional square-pyramid model, it is believed that $\df=\log_2 5$.  In both cases $\df<d$.)
The free energy gain on relaxing such an object is purely
entropic (the idea is that the initial state is localised in a single metastable minimum while the final state can choose from many similar states).  The configurational
entropy per site in the TPM is comparable with the total entropy, which scales as $s\sim (J/T) \ee^{-J/T}$.  Considering the growing droplet we therefore
estimate
\begin{equation}
\Delta F \approx J \log R - s R^{\df} .
\label{equ:lnt}
\end{equation}
This free energy barrier is maximal at 
$R^* \sim (J/s)^{1/\df}$.  Substituting for $s$, the barrier height is therefore $F^* \sim J^2/(T\df)$, leading to a relaxation time that scales as 
\begin{equation}
\log \tau \sim J^2/(T^2\df).
\label{equ:log-tau}
\end{equation}
This result coincides  
with the relaxation-time scaling for the TPM that is predicted 
and observed in numerics~\cite{Garrahan02,Jack05caging}, subject to numerical prefactors
in (\ref{equ:log-tau}) which are rather hard to establish, both in numerics~\cite{Jack05caging} and analytically (consider for example the simpler case of the 
East model~\cite{evans99,gst}).  We note that the length scale $R^*$ obtained from this argument is also of the same order as
the four-point correlation length at equilibrium, and the cavity point-to-set length, both of which scale as $(\ee^{-J/T})^{-1/\df}$: see Ref.~\onlinecite{Jack05caging}.

The resulting picture is that the interfacial costs for nucleation of relaxation of localised droplets can be understood in terms of a crossover formula,
$\Delta F_{\rm int} \sim J \log R + \gamma R^{d-1}$, with the logarithmic term being relevant for the relatively small droplets that control
 equilibrium relaxation, while the surface tension term ($\gamma R^{d-1}$) is relevant for large droplets, such as those found in nucleation close to first-order
phase boundaries.  Similarly, the bulk free energy gain from a droplet of size $R$ can be approximated as $\Delta F_{\rm bulk} \sim s R^{\df} + \Delta \mu R^d$
where again the first term is relevant for smaller droplets and equilibrium relaxation, and the second term applies to larger droplets, as found in nucleation and growth.

Of course, these arguments are based on several conjectures: it would be interesting to test them using further numerical studies.  However, they do seem to offer a coherent picture of the TPM dynamics and of its static many-body correlations (at least at the level of point-to-set).  In general, the idea that
nucleation-and-growth of relatively small droplets might occur with a non-classical free energy such as (\ref{equ:lnt}) follows the arguments in 
Ref.~\onlinecite{Bouchaud2004}, but with the additional
generalisation that even the bulk term might not scale as $R^d$.

\section{Outlook\label{sec:outlook}}

We have used the TPM to illustrate how phase transitions that occur in systems of coupled replicas can be used to rationalise the experimental behaviour of ultrastable glasses, constructing a direct connection with 
nucleation-and-growth dynamics.  
This provides a theoretical explanation for the compressed 
exponential Avrami kinetics and the giant length scales 
that are observed in experiments.

The TPM is a schematic model and does not describe the experimental system in detail, but these results show how predictions based on phase transitions and universal behaviour can be useful in practical settings.  The TPM combines facilitated dynamics of point-like excitations with static many-body spin correlations that can be long-ranged and lead to signficant amorphous order.  By combining these two ingredients, the model can capture many qualitative features of glass-forming systems, including non-trivial aging behaviour, dynamical heterogeneity, and both static and dynamic phase transitions. 

More generally, the present results should serve as useful guides 
to interpret future work dealing with the dynamics of stable glasses. 
In particular, our approach suggests that spatially resolved analsyis 
of the melting dynamics of {\it in-silico} stable glasses, or experimental 
materials would be very valuable in validating the present picture. 
Equilibrium dynamic heterogeneity is so short-ranged that 
direct measurements of dynamic correlation lengthscales remain scarce
for molecular liquids. We suggest that direct measurements 
of the non-equilibrium lengthscales discussed here could be much easier, 
as these lengthscales may be larger by orders of magnitude, and potentially
more easily accessible to experimental work.

\begin{acknowledgments}
We thank M. D. Ediger for helpful discussions.
The research leading to these results has
received funding from the European Research Council
under the European Union’s Seventh Framework
Programme (FP7/2007-2013)/ERC Grant Agreement
No. 306845.
\end{acknowledgments}


\end{document}